%% file: ijcai24.tex
\documentclass{article}
\usepackage{ijcai24}

\usepackage{times}
\usepackage{soul}
\usepackage{url}
\usepackage[hidelinks]{hyperref}
\usepackage[utf8]{inputenc}
\usepackage[small]{caption}
\usepackage{graphicx}
\usepackage{amsmath}
\usepackage{amsthm}
\usepackage{amsfonts}
\usepackage{booktabs}
\usepackage{algorithm}
\usepackage{algorithmic}
\usepackage[switch]{lineno}
\usepackage{dsfont}
\usepackage{subfig}
\usepackage{color}
\usepackage{multirow}

\newtheorem{lemma}{Lemma}
\newtheorem*{remark}{Remark}

\title{Probabilistically Robust Watermarking of Neural Networks}

\author{
Mikhail Pautov$^{1,2,3}$
\and
Nikita Bogdanov$^2$\and
Stanislav Pyatkin$^2$\and \\ 
Oleg Rogov$^{1,2}$\And 
Ivan Oseledets$^{1,2}$\\
\affiliations
$^1$Artificial Intelligence Research Institute, Moscow, Russia\\
$^2$Skolkovo Institute of Science and Technology, Moscow, Russia\\
$^3$ISP RAS Research Center for Trusted Artificial Intelligence, Moscow, Russia\\
\emails
\{mikhail.pautov, nikita.bogdanov, stanislav.pyatkin\}@skoltech.ru,
\{rogov, oseledets\}@airi.net
}

\begin{document}

\maketitle

\begin{abstract}
    As deep learning (DL) models are widely and effectively used in Machine Learning as a Service (MLaaS) platforms, there is a rapidly growing interest in DL watermarking techniques that can be used to confirm the ownership of a particular model. Unfortunately, these methods usually produce watermarks susceptible to model stealing attacks. In our research, we introduce a novel trigger set-based watermarking approach that demonstrates resilience against functionality stealing attacks, particularly those involving extraction and distillation. Our approach does not require additional model training and can be applied to any model architecture. The key idea of our method is to compute the trigger set, which is transferable between the source model and the set of proxy models with a high probability. In our experimental study, we show that if the probability of the set being transferable is reasonably high, it can be effectively used for ownership verification of the stolen model. We evaluate our method on multiple benchmarks and show that our approach outperforms current state-of-the-art watermarking techniques in all considered experimental setups.

    



    
\end{abstract}

\section{Introduction}

Deep learning models achieved tremendous success in practical problems from different areas, such as computer vision \cite{he2016deep,dosovitskiy2021an}, natural language processing \cite{vaswani2017attention,brown2020language} and multimodal learning \cite{tang2022mmt}. They are used in medical diagnostics \cite{he2023transformers,goncharov2021ct}, deployed in autonomous vehicles \cite{huang2022survey,parekh2022review} and embedded in AI-as-a-service settings \cite{buhalis2022voice,liu2023summary}. Unfortunately, the development, training, and production of these models nowadays come at a high cost due to the availability and quality of the training data, the large size of the models, and, hence, the necessity of cloud computing and data storage platforms. This motivates the owners to prevent  third parties from obtaining an illegal copy of their models, acquiring the powerful tools without spending much time and money on development and training. 


\begin{figure*}[tb]
    \centering
    \includegraphics[width=1.0\textwidth]{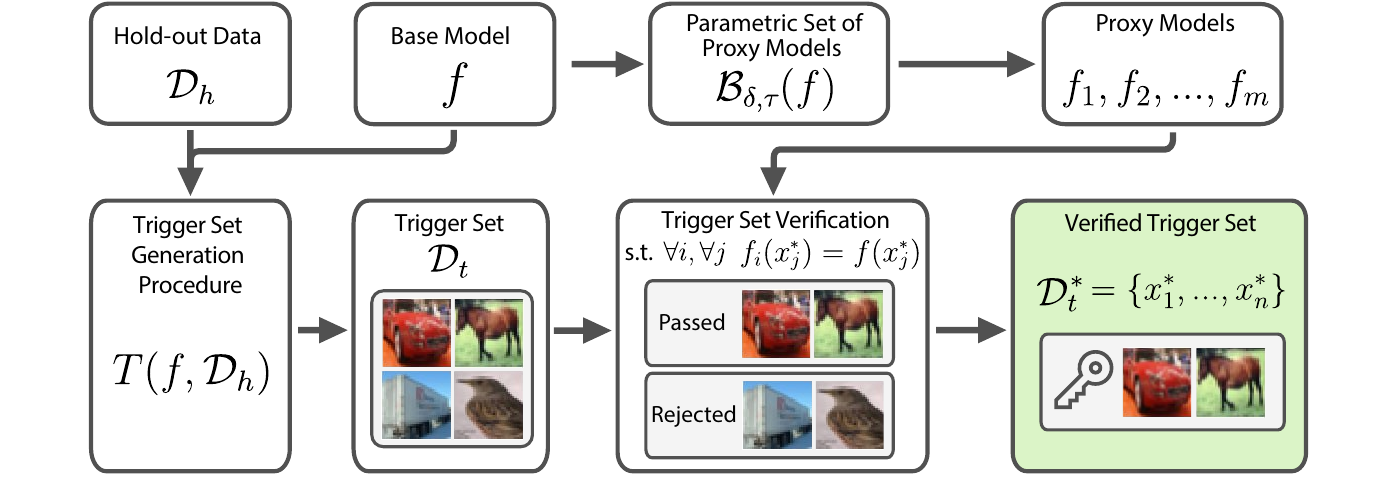}
    \caption{The illustration of the proposed pipeline for the trigger set generation and verification. Given the source model $f$ and the hold-out data $\mathcal{D}_h$, we initialize the parametric set of proxy models $\mathcal{B}_{\delta, \tau}(f)$ introduced in Equation~\eqref{eq:proxy_set} and sample $m$ proxy models $f_1,\dots,f_m$ from this set. Then, given the procedure of trigger set generation $T = T(f, \mathcal{D}_h)$, we compute the trigger set candidates $\mathcal{D}_t$. The samples from the candidate set $\mathcal{D}_t$ that are verified by the proxy models $f_1,\dots, f_m$ are included in the verified trigger set $\mathcal{D}^*_t.$ The procedure is executed until the verified trigger set of size $n$ is collected.}
    \label{fig:big_teaser}
\end{figure*}

Among the methods to protect the copyright of digital assets, digital watermarking techniques \cite{hartung1999multimedia} are the most widely used. To determine the copyright violation, the intellectual property owner embeds the special information in the product, for example, adding an imperceptible pattern or digital signature to the image or within the program's source code. If the violation is suspected, this information may be extracted from the intellectual property, confirming the illegal obtaining of the latter. In recent years, watermarking techniques have been adapted to protect the ownership of deep learning models embedded in the black-box manner. To do so, the owner of the model may prepare the special (trigger) set of points the source model should have the specific predictions on: the more similar the predictions of a suspicious on this set to the prespecified ones, the more likely it is that the source model has been compromised \cite{zhang2018protecting,adi2018turning,bansal2022certified,kim2023margin}. 

In practice, watermarks are not resistant to attacks that are aimed to steal the model's functionality. In particular, distillation attacks, fine-tuning, and regularization of models tend to affect the transferability of trigger sets \cite{shafieinejad2021robustness}. Thus, researchers are motivated to explore the robustness of watermarks to stealing attacks.

In this paper, we propose a novel framework that enhances the resistance of trigger set-based watermarks to stealing attacks. Given the source model $f$, we construct a parametric set $\mathcal{B}_{\delta, \tau}(f)$ of proxy models which imitate the set of surrogate (or stolen) copies of $f$. We assume that, given the parametric set of proxy models, there exist input data points $\mathcal{S}(f, \delta, \tau)$ that all the models assign to the same class. If the behavior of the source model on these points is prespecified, they are treated as good trigger points for ownership verification.  In our method, we ensure that all the proxy models assign a particular sample from the trigger set to some specific class by comparing the predictions of $m$ randomly sampled proxy models (see Figure \ref{fig:big_teaser}).

We summarize the contributions of this work as follows:
\begin{itemize}
    \item 
    We introduce a novel probabilistic approach for enhancing trigger-set-based watermarking methods' robustness against stealing attacks.  Our approach can be applied to enhance the robustness of any trigger set-based watermarking technique, which makes it universal.
    \item    We analyze the probability that a given trigger set is transferable to the set of proxy models that mimic the stolen models.
    \item We experimentally show that, even if the stolen model does not belong to the set of proxy models, the trigger set is still transferable to the stolen model. 
    \item We evaluate our approach on multiple benchmarks and show that it outperforms current methods in all considered experimental setups.

\end{itemize}

\section{Related Work}
\label{s:related}

The process of DNN watermarking serves the purpose of safeguarding intellectual property by encoding a distinctive pattern and employing it for the purpose of asserting ownership \cite{ahmed:ieee-steganography,li2021survey}. 

Certain watermarking techniques insert digital media watermarks into the initial training data to create a trigger dataset for the model. For example, \cite{Guo.2018Watermarking} generates an $n$-bit signature representing the model owner and embeds it into the training data to create the trigger dataset. The authors ensure that the altered images in the trigger dataset receive unique labels different from the original data points.

In \cite{zhang2018protecting}, the authors  proposed algorithms for watermarking neural networks used in image classification, along with remote black-box verification methods. One technique involves adding meaningful content alongside the original training data to create a watermark. For example, they embed a unique string, like a company name, into an image from the training set during the prediction process, assigning a different label to the modified sample. Alternatively, noise can be added to the original training data as part of the watermarking process.

A similar approach suggested in \cite{Li.2019How} features blending the regular data samples with distinctive ``logo'' elements and training the model to categorize them under a specific label. To maintain similarity with the original samples, they use an autoencoder, and its discriminator is trained to distinguish between benign training samples and watermark-containing trigger samples.


In contrast, there are studies on certain disadvantages of trigger set-based watermarking methods. 

First, this category of methods has a limitation tied to the maximum number of backdoors that can be integrated into a neural network. There are works showing that the large number of watermarked samples in the training set leads to notable performance  degradation of the source model \cite{jia2021entangled,kim2023margin}.

Secondly, watermarking schemes lacking a verifiable connection between the watermark and the legitimate model owner create an opportunity for attackers to counterfeit the watermark \cite{adi2018turning,Guo.2018Watermarking}.
Lastly, the use of adversarial examples for trigger set-based watermarking \cite{lemerrer:stiching} or fingerprinting \cite{lukas:iclr-fingerprint,zhao:finger-auth} has significant drawbacks, including insufficient transferability to the surrogate models \cite{kim2023margin}, potential vulnerability to knowledge distillation attacks \cite{hinton2015distilling}, fine-tuning and retraining. 


\section{Problem Statement}
\label{s:problem}

\subsection{Trigger Set-Based Digital Watermarking}

In our work, we consider classification problem with $K$ classes. Namely, given the dataset $\mathcal{D} = \{(x_i, y_i)\}_{i=1}^N$, where $x_i \in \mathbb{R}^d$ and $y_i \in [1,\dots, K],$ we train the source model $f$ to minimize the empirical risk 

\begin{equation}
\label{eq:common_loss}
    L(\mathcal{D}) = \frac{1}{N} \sum_{i=1}^N l(f(x_i), y_i),
\end{equation}
where $l(\cdot,\cdot)$ is the cross-entropy loss.

If the source model performs well, an adversary can try to steal its functionality. Namely, one can train surrogate  model $f^{*}$ on surrogate dataset $\hat{\mathcal{D}}$ that aims to imitate the outputs of the base model. In general, the architectures of the source model and the  surrogate one do not have to be the same. It is equivalently not required that the initial dataset is known, which makes it possible to steal the models in the black-box manner  \cite{jia2021entangled,kim2023margin}.    

To detect theft, the owner of the source model can apply trigger set-based watermarking.  For example, the subset $\mathcal{D}_s = \{(x_{i_k}, y_{i_k})\}_{k=1}^n$ of the initial dataset $\mathcal{D}$ undergoes label flipping: each label $y_{i_k}$ is replaced with another label $y^{'}_{i_k} \ne y_{i_k}$ yielding the trigger set $\mathcal{D}_t = \{(x_{i_k}, y^{'}_{i_k})\}_{k=1}^n$. Then, the source model is trained to minimize the empirical risk on the changed dataset $\mathcal{D} := (\mathcal{D} \setminus \mathcal{D}_s) \cup \mathcal{D}_t.$

If the performance of suspicious model $f^{*}$ on the trigger set $\mathcal{D}_t$ is similar to the one of the source model $f,$ it is claimed that the source model is stolen. 

Unfortunately, trigger set-based watermarking approaches have two significant drawbacks. Firstly, the size $n$ of the trigger set has to be small to not cause a notable performance decrease. On the other hand, $n$ should be sufficiently large so that the similarity in the behavior of the source model and the stolen model is statistically significant.  Secondly, trigger sets tend to be barely transferable between the source model and the stolen model: a sample from a trigger set is pushed away from the samples of the same class closer to the decision boundary of the surrogate model \cite{kim2023margin}. 

In this work, we propose a simple and effective  approach to generate the trigger set $\mathcal{D}_t$, which is transferable between the source model and its surrogate copies obtained by model stealing attacks.

\subsection{Model Stealing Attacks}

In our work, we assume that an adversary attempts to steal the source model by applying knowledge distillation\footnote{We include additional experiments with other stealing attacks in the supplementary material.} \cite{hinton2015distilling,stealing-ijcai:tramer} considering it the strongest attack for the watermark removal \cite{kim2023margin}. 

In particular, a sample $\hat{x}_i$ from the surrogate dataset $\hat{\mathcal{D}}$ is passed to the source model $f$ to obtain its prediction $f(\hat{x}_i)$. Then, the stealing is performed by training a new model $f^{*}$ by minimizing the divergence between its predictions and the source model's predictions on the dataset $\hat{\mathcal{D}}$:

\begin{equation}
    \label{eq:stealingloss}
    L_\text{ext}(\hat{\mathcal{D}}) = \frac{1}{|\hat{\mathcal{D}}|} \sum_{\hat{x_i} \in \hat{\mathcal{D}}} D_\text{KL} (f(\hat{x}_i), f^{*}(\hat{x}_i)),
\end{equation}
where $D_\text{KL}$ is Kullback–Leibler divergence. 

We consider both soft-label and hard-label attacks, i.e., $f^{*}(\hat{x}_i)$ can be either the predicted class or vector of class probabilities.


\section{Method}
\label{s:method}


\subsection{Computing the Trigger Set}
\label{ss:trigger_set_proc}
In our approach, we exploit the procedure of computing the candidates for the trigger set $\mathcal{D}_t$ as convex combinations of the pairs of points from the hold-out dataset \cite{DBLP:conf/iclr/ZhangCDL18}. Namely, suppose the source model $f$ is trained on the dataset $\mathcal{D}$.  Then, given the hold-out test data $\mathcal{D}_h: \mathcal{D}_h \cap \mathcal{D} = \emptyset,$ we uniformly sample a pair of points $(x_{i_1}, y_{i_1}), (x_{i_2}, y_{i_2})$ from different classes $y_{i_1}$ and $y_{i_2} \ne y_{i_1}$, and compute the convex combination of $x_{i_1}$ and $x_{i_2}$ in the form 

\begin{equation}
    \label{eq:convex}
    x_i^{*} = \lambda x_{i_1} + (1-\lambda)x_{i_2},
\end{equation}
where $\lambda \sim \mathcal{U}(0,1)$. To assure the unexpected behaviour of the model $f$ on the trigger set candidate, we accept $x_i^{*}$ as the candidate only if the source model predicts $x_i^{*}$ as the sample from some other class $y_i^{*}: y_i^{*} \ne y_{i_1}$ and  $y_i^{*} \ne y_{i_2}.$  The procedure of computing the candidates for the trigger set is presented in Algorithm~\ref{alg:trigger_candidate}. We execute Algorithm~\ref{alg:trigger_candidate} until the candidate set $\mathcal{D}_t = \{(x^{*}_i, y^{*}_i)\}_{i=1}^n$ is computed. Note that the described procedure requires no additional training of the source model.

\begin{algorithm}[tb]
    \caption{Trigger set candidate}
    \label{alg:trigger_candidate}
    \textbf{Input}: Hold-out dataset $\mathcal{D}_h$, source model $f$ \\
    \textbf{Output}: Trigger set candidate $(x^*, y^*)$
    \begin{algorithmic}[1] 
        \WHILE{True}
        \STATE Sample $(x_{1}, y_{1}), (x_{2}, y_{2}) \sim \mathcal{U}(\mathcal{D}_h)$ \label{alg1:sample}
        \IF {$y_{1} \ne y_{2}$}
        \STATE Sample $\lambda \sim \mathcal{U}(0,1)$
        \STATE $x^{*} = \lambda x_{1} + (1-\lambda) x_{2}$
        \STATE $y^{*} = f(x^{*})$

        \IF {$y^{*} \ne y_{1}\ \text{and}\ y^{*} \ne y_{2}$}
        \STATE \textbf{return} $(x^{*}, y^{*})$
        \ENDIF
        \ENDIF
        \ENDWHILE
    \end{algorithmic}
\end{algorithm}

\subsection{Verification of the Trigger Set}

The core idea of our method is to ensure that the trigger set is transferable to the stolen models or, in other words, to verify that the predictions of a stolen trigger set are similar to the ones of the source model. To do so, we introduce the parametric set of models $\mathcal{B}_{\delta, \tau}(f)$  that mimics the set of stolen models.  In our experiments, we mainly consider the case when the architecture of the source model is known to a potential adversary. Hence, this parametric set consists of proxy models $f^{'}$ of the same architecture as the source $f$ that perform reasonably well on the training dataset $\mathcal{D}$. Namely, if $\theta(f)$ is the flattened vector of weights of the model $f$, the parametric set  $\mathcal{B}_{\delta, \tau}(f)$ is defined as follows:

\begin{align}
    \label{eq:proxy_set}
    \begin{split}
    \mathcal{B}_{\delta, \tau}(f) = &\{f^{'}: \|\theta(f^{'}) - \theta(f)\|_2 \le \delta \ \text{and} \\ &|\text{acc}(\mathcal{D},f^{'}) - \text{acc}(\mathcal{D},f)| \le \tau \},
    \end{split}
\end{align}
where $\text{acc}(\mathcal{D},f)$ is the accuracy of model $f$ on the dataset $\mathcal{D}$, $\delta$ is the weights threshold and $\tau$ is the performance threshold.

To verify the transferability of the trigger set,  we sample $m$ proxy models $f_1, \dots, f_m$ from $\mathcal{B}_{\delta, \tau}(f).$ Then, we check if all $m$ proxy models assign the same class label to the samples of the trigger set as the source model $f.$ The procedure of trigger set verification is presented in Algorithm~\ref{alg:trigger_set_ver}. The method of computing the verified trigger set is illustrated in Figure~\ref{fig:big_teaser}. 
\begin{remark}\label{eq:sample_proxy}
    In our experiments, we sample proxy model $f_i \sim \mathcal{B}_{\delta, \tau}(f)$ by generating a noise vector $\Delta_i \sim \mathcal{N}(0, \sigma^2I)$ for some $\sigma^2>0$ and assuring $\|\Delta_i\|_2 \le \delta.$ Then, the vector of weights of proxy model $f_i$ is computed as $\theta(f_i) \gets \theta(f) + \Delta_i.$
\end{remark}

\begin{algorithm}[tb]
    \caption{Trigger set verification}
    \label{alg:trigger_set_ver}
    \textbf{Input}: Hold-out dataset $\mathcal{D}_h$, source model $f$, weights threshold $\delta$, performance threshold $\tau$, verified trigger set size $n$\\
    \textbf{Output}: Verified trigger set $\mathcal{D}^{*}_t$
    \begin{algorithmic}[1] 
        \STATE Initialize $\mathcal{B}_{\delta, \tau}(f)$
        \STATE Sample $f_1,\dots, f_m \sim \mathcal{B}_{\delta, \tau}(f)$
        \STATE Let $i=0$
        \STATE Let $\mathcal{D}^*_t = \emptyset$
        \WHILE{$i < n$}
        \STATE $(x_i^*, y_i^*) \gets \text{TriggerSetCandidate}(\mathcal{D}_h, f)$ \label{alg2:candidate}
        \IF {$f_1(x_i^*) = f_2(x_i^*) = \dots = f_m(x_i^*) = y_i^*$}
        \STATE $\mathcal{D}^*_t \gets \mathcal{D}^*_t \cup \{(x_i^*, y_i^*)\}$
        \STATE $i \gets i+1$
        \ENDIF
        \ENDWHILE
        \STATE \textbf{return} $\mathcal{D}^*_t$
    \end{algorithmic}
\end{algorithm}

\section{Experiments}
\label{s:experiments}


\subsection{Setup of Experiments}

\subsubsection{Datasets and Training} In our experiments, we use CIFAR-10 and CIFAR-100 \cite{krizhevsky2009learning} as training datasets for our source model $f.$ For the purpose of comparison, as the source model, we use ResNet34 \cite{he2016deep}, which is trained for $100$ epochs to achieve high classification accuracy (namely, 91.0\% for CIFAR-10 and 66.7\% for CIFAR-100). We used SGD optimizer with learning rate of $0.1$, weight decay of $0.5 \times 10^{-3}$  and momentum of $0.9.$ 

\subsubsection{Parametric Set of Proxy Models}
\label{sub:parametric}
Once the source model is trained, we initialize a  parametric set of proxy models $\mathcal{B}_{\delta, \tau}(f).$ In our experiments, we vary the parameters of the proxy models set to achieve better trigger set accuracy of our approach. Namely, parameter $\delta$ was varied in the range $[0.5, 40]$ and $\tau$ was chosen from the set $\{0.1, 0.2, 1.0\}.$ We tested different number of proxy models sampled from $\mathcal{B}_{\delta, \tau}(f)$ for verification. Namely, parameter $m$ was chosen from the set $\{1,2,4,8,16,32,64,128,256\}.$ 

\subsubsection{Model Stealing Attacks}
Following the other works \cite{jia2021entangled,kim2023margin}, we perform functionality stealing attack by training the surrogate model $f^{*}$ in the following three settings:

\begin{itemize}
    \item Soft-label attack. In this setting, the training dataset $\mathcal{D}$ is known, and, given input $x$, the output $f(x)$ of the source model is a vector of class probabilities. The surrogate model $f^{*}$ is trained to minimize the functional from Eq.~\eqref{eq:stealingloss}.
    \item Hard-label attack. In this setting, the training dataset $\mathcal{D}$ is known, and, given input $x$, the output $f(x)$ of the source model is the class label assigned by $f$ to input $x$. This setting corresponds to the training of the surrogate model on the dataset $\hat{\mathcal{D}} = \{x_i, f(x_i)\}_{i=1}^N.$
    \item Regularization with ground truth label. In \cite{kim2023margin}, it was proposed to train the surrogate model by minimizing the empirical loss on the training dataset $\mathcal{D}$ and the KL-divergence  between the outputs of the source model and surrogate model simultaneously. This setting corresponds to the minimization of the convex combination of the losses from Eq.~\eqref{eq:common_loss} and Eq.~\eqref{eq:stealingloss} in the form 
    \begin{equation}
        \label{eq:rgt_loss}
        L_\text{RGT}(\mathcal{D}, \hat{\mathcal{D}}, \gamma) = \gamma L_\text{ext}(\hat{\mathcal{D}}) + (1-\gamma) L(\mathcal{D}),
    \end{equation}
    where $\gamma \in [0,1]$ is the regularization coefficient. In our experiments, this is the strongest functionality stealing attack. 
\end{itemize}

\subsubsection{Concurrent Works}
We evaluate our approach against the following methods. 

\begin{itemize}
    \item Entangled Watermark Embedding (EWE). In~\cite{jia2021entangled}, it was proposed to embed watermarks by forcing the source model to entangle representations for legitimate task data and watermarks.  
    \item Randomized Smoothing for Watermarks (RS). In~\cite{bansal2022certified}, randomized smoothing is applied to the parameters of the source model, yielding guarantees that watermarks can not be removed by a small change in the model's parameters.
    \item Margin-based Watermarking (MB). In~\cite{kim2023margin},  it was proposed to train the surrogate model by pushing the decision boundary away from the samples from the trigger set so that their predicted
    labels can not change without compromising the
    accuracy  of the source model.
\end{itemize}
It is worth mentioning that all the baselines we compare our approach against either require modification of the training procedure of the source model or make its inference computationally expensive.   

\subsubsection{Evaluation Protocol}
Once the verified trigger set $\mathcal{D}^{*}_t = \{(x^{*}_i, y^{*}_i)\}_{i=1}^n$ is collected and surrogate model $f^{*}$ is obtained, we measure the accuracy 

\begin{equation}
    \label{eq:trigger_set_acc}
    \text{acc}(\mathcal{D}_t^{*}, f^{*}) = \frac{1}{|\mathcal{D}_t^{*}|} \sum_{(x^{*}_i, y^{*}_i) \in \mathcal{D}_t^{*}} \mathds{1}\left(f^{*}(x^{*}_i) = y^{*}_i \right)
\end{equation}
of $f^{*}$ on  $\mathcal{D}^{*}_t$ to evaluate the effectiveness of our watermarking approach. 

\begin{remark}
    Later, to compare our approach with concurrent works, we denote the trigger set as $\mathcal{D}^{*}$ to emphasize the differences in trigger set collection procedures.
\end{remark}

\subsubsection{Parameters of Experiments}
Unless stated otherwise, we use the following values of hyperparameters in our experiments: the size of the verified trigger set $n$ is set to be $n=100$ for consistency with the concurrent works,
confidence level $\alpha$ for Clopper-Pearson test from Eq.~\eqref{eq:cp_int} is set to be $\alpha=0.05$. In our experiments, we found that better transferability of the verified trigger set is achieved when no constraint on the  performance of the proxy models is applied, so the performance threshold parameter is set to be $\tau=1.0$.

\subsection{Results of Experiments}

\subsubsection{Knowledge Required for Model Stealing}
In our experiments, we assume either the architecture of the source model or its training dataset is \emph{known} to a potential adversary.
In Table~\ref{tab:main_table}, we present the results for the most aggressive stealing setting, i.e., the one where the adversary is aware of both the architecture and the training data of the source model. In Table~\ref{tab:table_additional}, we report the results for the setting where either architecture or the training dataset is \emph{unknown} to the adversary. Namely, following \cite{kim2023margin}, we (i) perform a model stealing attack using the surrogate dataset SVHN~\cite{netzer2011reading} and (ii) perform stealing attack by replacing the architecture of the surrogate model by VGG11~\cite{DBLP:journals/corr/SimonyanZ14a}.

For each experiment, we train a single instance of the source model $f$ and perform $N_\text{st}=10$ independent model stealing attacks. For our approach and concurrent works, we report the accuracy of the source model $f$ and the surrogate models $f^{*}$ on training dataset $\mathcal{D}$ and trigger set $\mathcal{D}^{*}$.  It is notable that our approach not only outperforms the baselines in terms of trigger set accuracy but also yields the source model with higher accuracy. 

\input{FormattingGuidelines-IJCAI-24/table_with_diffs}
\input{FormattingGuidelines-IJCAI-24/table1}
\input{FormattingGuidelines-IJCAI-24/table_additional_exp}
\input{FormattingGuidelines-IJCAI-24/table_with_parameters}

\subsubsection{Hyperparameters Tuning}
It is important to mention that the parameters of proxy set $\mathcal{B}_{\delta, \tau}(f)$  affect not only the transferability of the trigger set  but also the computation time needed to collect the trigger set. Indeed, the larger the value of  $\delta$ is, the more proxy models are used for verification of the trigger set, the more often the procedure from Algorithm~\ref{alg:trigger_set_ver} rejects trigger set candidates. To find the trade-off between the accuracy of the surrogate model on the trigger set and the computation time, we perform hyperparameter tuning. We tune parameters $m$ and $\delta$ according to Section \ref{sub:parametric}. 

According to parameters tuning, we choose $m=64$ and $\delta=40.0$ as the default parameters of the proxy set. In Table~\ref{tab:parameters_table}, we report the values of parameters we used in each experiment. It is noteworthy that tuning only parameters of the proxy set $\mathcal{B}_{\delta, \tau}(f)$ allows our method to surpass existing approaches by a notable margin.

\section{Discussions}

\subsection{Transferability of the Verified Trigger Set}

In our approach, we assume that all the models from the parametric set $\mathcal{B}_{\delta, \tau}(f)$ are agreed in predictions on data samples from unknown \emph{common set} $\mathcal{S}(f, \delta, \tau).$ In other words, if $f(x)$ is the class assigned by model $f$ to sample $x,$ the set $\mathcal{S}(f, \delta, \tau)$ is defined as follows:

\begin{equation}
    \label{eq:common_points}
    \mathcal{S}(f, \delta, \tau) = \{x: f(x) = f^{'}(x) \ \forall f^{'} \in \mathcal{B}_{\delta, \tau}(f)\}.
\end{equation}
If the stolen model belongs to the set of proxy models $\mathcal{B}_{\delta, \tau}(f)$, a trigger set build-up from points from common set $\mathcal{S}(f, \delta, \tau)$ would be a good candidate for ownership verification: by design, the predictions of the source model and the stolen model would be identical on such a set. 

Since it is impossible to guarantee that a certain data point belongs to the common set $\mathcal{S}(f, \delta, \tau),$ we perform the screening of the input space for the candidates to belong to $\mathcal{S}(f, \delta, \tau).$ 

Namely, given a candidate $x$, we check the agreement in predictions of  $m$ randomly sampled proxy models $f_1, \dots, f_m$ from $\mathcal{B}_{\delta, \tau}(f)$ and accept $x$ as the potential member of  $\mathcal{S}(f, \delta, \tau)$ only if all $m$ models have the same prediction. One can think of the selection process of such points as tossing a coin: checking the predictions of $m$ proxy models represents $m$ coin tosses. The input data points represent unfair coins, i.e., those with different probabilities of landing on heads and tails. If the input point $x$ and the index of proxy model $i$ is fixed, such an experiment $A_i = A_i(x)$ is a  Bernoulli trial:  

\begin{equation}
    \label{eq:bernoulli}
    A_i(x) = 
    \begin{cases}
      1 & \text{with probability } p(x), \\
      0 & \text{with probability } 1-p(x).
    \end{cases} 
\end{equation}
Let the success of the Bernoulli trial from Eq.~\eqref{eq:bernoulli} correspond to the agreement in predictions of the source model $f$ and $i-$th proxy model $f_i$. Thus, the screening reduces to the search of input points  with the highest probability $p(x)$.

In our experiments, we estimate the parameter $p(x)$ of the corresponding random variable by observing the results of $m$ experiments $A_1(x), \dots, A_m(x)$. We use interval estimation for $p(x)$ in the form of Clopper-Pearson
test \cite{clopper1934use} that returns one-sided $(1-\alpha)$ confidence interval for $p(x)$:

\begin{equation}
    \label{eq:cp_int}
    \mathbb{P}\left(p(x) \ge B\left(\frac{\alpha}{2}, t, m-t+1\right)\right) \ge 1-\alpha.
\end{equation}
In Eq.~\ref{eq:cp_int}, $\hat{p}(x) = B\left(\frac{\alpha}{2}, m, 1\right)$ is the quantile from the Beta distribution and the number of successes $t=m$. The following Lemma gives probabilistic guarantees on the transferability of predictions on the verified trigger set from the source model to a proxy model from the set $\mathcal{B}_{\delta, \tau}(f).$


\begin{lemma}
Given the sampling procedure for proxy models from Section \ref{eq:sample_proxy}, the confidence level $\alpha$ from Eq. \eqref{eq:cp_int}, with probability at least $\phi = (1-\alpha)^n,$ the expectation of accuracy of the proxy model $f_i \sim \mathcal{B}_{\delta, \tau}(f)$ on the verified trigger set  $\mathcal{D}^{*}_t$ of size $n$ is at least $\text{acc}(\mathcal{D}^{*}_t, f^{*}) = \hat{p}(x).$

\begin{proof}
    Note that with probability at least  $\phi = (1-\alpha)^n$ the interval estimations for $p(x)$ from Eq. \eqref{eq:cp_int} hold for all the $n$ samples from $\mathcal{D}^{*}_t.$ Fixing the proxy model $f_i \sim  \mathcal{B}_{\delta, \tau}(f)$, we can compute its accuracy  $\text{acc}(\mathcal{D}^{*}_t, f^{*})$ on the verified trigger set:

    \begin{equation}
    \label{eq:expected_acc}
        \text{acc}(\mathcal{D}^{*}_t, f^{*}) = \frac{1}{n} \sum_{(x_j,y_j) \in \mathcal{D}^{*}_t} A_i(x_j), 
    \end{equation}
    where $A_i(x_j)$ is in the form from Eq. \eqref{eq:bernoulli}. 
    Taking expectation of Eq. \eqref{eq:expected_acc} yields
    \begin{equation}
        \begin{split}
        & \mathbb{E} \left(\text{acc}(\mathcal{D}^{*}_t, f^{*})\right)  =  \frac{1}{n} \sum_{(x_j,y_j) \in \mathcal{D}^{*}_t} \mathbb{E}(A_i(x_j)) \\
        & = \frac{1}{n} \sum_{(x_j,y_j) \in \mathcal{D}^{*}_t} p(x_j) \ge \frac{1}{n} \sum_{(x_j,y_j) \in \mathcal{D}^{*}_t} \hat{p}(x_j) \equiv \hat{p}(x).
    \end{split}
    \end{equation}
\end{proof}
\end{lemma}

\begin{remark}
    If the distributions of the proxy models and the surrogate models on $\mathcal{B}_{\delta, \tau}(f)$ are the same, the Lemma above yields probabilistic guarantees on the transferability of the predictions on the trigger set to surrogate models. 
\end{remark}

We treat $\hat{p}(x)$ as the lower bound of $p(x)$; the higher the value of $\hat{p}(x)$, the higher the probability that a certain sample $x$ belongs to the common set $\mathcal{S}(f, \delta, \tau)$, given the finite number of proxy models used for verification.

In the majority of our experiments, we use $m=64$ proxy models for the verification of the trigger set. It yields a uniform lower bound $\hat{p}(x) \ge 0.9$ from Eq.~\ref{eq:cp_int} for all the samples $x$ that are included in the verified trigger set.
It is notable that such a moderate lower bound in practice leads to high transferability of the trigger set to the surrogate models. 

However, according to our experimental settings, surrogate models $f^{*}$ \emph{do not have to} belong to the proxy set $\mathcal{B}_{\delta, \tau}(f)$ due to plausible difference in architectures or large difference in weights. Hence, it is not guaranteed that surrogate models have the same common set $\mathcal{S}(f,\delta,\tau)$ as proxy ones.   In Table~\ref{tab:table_diffs}, we report the norm of the difference of models' parameters between the source model and surrogate models. It is noteworthy that the surrogate models do not belong to the proxy set $\mathcal{B}_{\delta, \tau}(f)$. Despite this, our approach yields trigger sets that, in practice, are transferable beyond the proxy set.

\subsection{Integrity of the Method}
\label{s:integrity_sec}
It should be mentioned that a watermarking approach not only should not affect the source model's performance and be robust to stealing attacks, it should also satisfy the property of integrity. In other words, it should not judge non-watermarked networks as watermarked ones.  In trigger set-based watermarking settings, checking if a model is stolen may be thought of as a detection problem with certain false positive and false negative rates. The first one corresponds to the probability that a benign model is detected as stolen, and the second one corresponds to the probability that a stolen model is not detected  as such. 

Assuming that a stolen model belongs to the parametric set of proxy models $\mathcal{B}_{\delta, \tau}(f)$,  it is possible to provide probabilistic guarantees that the one would be detected as stolen by our method. In contrast, it is, in general, nontrivial to guarantee that a benign model would not be detected as stolen. With our method, such guarantees may be provided under certain modifications of the verification procedure. Namely, one may assume that all the models that  belong to the compliment $\bar{\mathcal{B}}_{\delta, \tau}(f)$ of the set of proxy models $\mathcal{B}_{\delta, \tau}(f)$ \emph{are not stolen} ones. Then, the verification procedure may be adapted: given models $f_1,\dots, f_m \in \mathcal{B}_{\delta, \tau}(f)$ and models $\bar{f}_1,\dots, \bar{f}_m \in \bar{\mathcal{B}}_{\delta, \tau}(f)$, the sample $(x^{*}, y^{*})$ is verified iff: 

\begin{equation}
    \label{eq:new_verify}
    \begin{cases}
    & y^{*} = f_1(x^{*}) = \dots = f_m(x^{*}),  \    \\
    & y^{*} \ne \bar{f}_1(x^{*}), \\
    & \dots\\
    &  y^{*} \ne \bar{f}_m(x^{*}).
    \end{cases}
\end{equation}
In other words, it is also required that the models from $\bar{\mathcal{B}}_{\delta, \tau}(f)$ are \emph{not} agreed with the source model on the samples from trigger set. It is notable that such a verification procedure requires careful parameterization of the set of proxy models: underestimation of its parameters would lead to some stolen models not being included in it, and overestimation of its parameters may lead to the inclusion of the benign models. 

\subsubsection{Experiments on the Integrity of the Method}

To satisfy the integrity property, the method has to be able to distinguish between stolen models and independent (not stolen) models $g$. To check this property, we evaluate  our approach on independent models $g$, which were trained in several different setups: 

\begin{itemize}
    \item In the first setting, we train $N_i = 64$ models on random subsets of initial training dataset. Architectures of independent models are either ResNet34 or WideResNet28 \cite{ZagoruykoK16}. For each model, the training dataset is twice as small as the dataset of the source model. We report the results of this setting in Table \ref{tab:integrity1}.

    \item In the second setting, we evaluate models of different architectures trained on the whole CIFAR-10 dataset. We consider VGG \cite{DBLP:journals/corr/SimonyanZ14a}, ShuffleNetV2 \cite{ma2018shufflenet}, ResNet20, ResNet32, ResNet44, ResNet56 \cite{he2016deep}, RepVGG \cite{ding2021repvgg} and MobileNetV2 \cite{sandler2018mobilenetv2} architectures. We report the results of this setting in Table \ref{tab:integrity2}.

    \item In the third setting, we train $N_i=3$ models of different architectures on different dataset, $\mathcal{D}=$ SVHN \cite{netzer2011reading}. All the models were trained for $30$ epochs with SGD optimizer, learning rate of $0.1$, and weight decay of $10^{-4}$. We report the results of this setting in Table \ref{tab:integrity3}. 
\end{itemize}
It is expected that if the model is independent, then it is less similar to the source model and, hence, its accuracy on the trigger set is lower than that of the source model and stolen models. As the results, we report the accuracy of independent models on respective training datasets $\mathcal{D}$ and accuracy on watermarks $\mathcal{D}_t^{*}$.

Notably, the independent models from the setups considered have different degrees of similarity with the source model.  Regardless of architecture, the model trained on the same dataset as the source model partially retains the behavior of the source model on the watermarks. However, trigger set accuracy is significantly lower than that of surrogate (stolen) models. Our experiments show that the least similar independent models are the ones trained on the different datasets, regardless of architecture.

The key advantage of our method is that treating the stealing attack as a random process, we obtain the probabilistic guarantees on the transferability of the predictions on the trigger set from the source model to the stolen one utilizing random (proxy) models from  $\mathcal{B}_{\tau, \delta}(f)$ and $\bar{\mathcal{B}}_{\tau, \delta}(f)$. To sum up, we want to outline that our model can catch the difference between the stolen model and independent models; the larger the dissimilarity between the independent model and the source model, the greater the difference in behavior on the trigger set.

\begin{table}[t]
    \centering
\begin{tabular}{cccc}
\toprule
 $g$ & acc($\mathcal{D}, f$) & acc($\mathcal{D}, g$) & acc($\mathcal{D}^{*}_t, g$) \\
\midrule
ResNet34 & \multirow{2}{*}{$90.6 \pm 0.4$} & $77.9 \pm 2.6$ & $55.7 \pm 7.2$ \\

WideResNet28 &  & $92.3 \pm 0.3$ & $53.9 \pm 5.0$ \\

\bottomrule
\end{tabular}
    \caption{Results of experiments on the integrity of the method, part $1$. The training dataset $\mathcal{D}$ is CIFAR-10.}
    \label{tab:integrity1}
\end{table}

\begin{table}[t]
    \centering
\begin{tabular}{cccc}
\toprule
 $g$ & acc($\mathcal{D}, f$) & acc($\mathcal{D}, g$) & acc($\mathcal{D}^{*}_t, g$) \\
\midrule
VGG & \multirow{5}{*}{$90.6 \pm 0.4$} & $93.7 \pm 0.5$ & $58.2 \pm 3.9$ \\
 
 ShuffleNetV2 & & $92.6 \pm 1.5$ & $57.3 \pm 2.7$\\
 
ResNet &  & $93.6 \pm 0.7$ & $53.9 \pm 5.0$ \\

RepVGG & & $94.8 \pm 0.3$ & $53.3 \pm 2.4$\\
MobileNetV2 & & $93.6 \pm 0.5$ & $57.0 \pm 2.2$\\

\bottomrule
\end{tabular}
    \caption{Results of experiments on the integrity of the method, part $2$. The training dataset $\mathcal{D}$ is CIFAR-10.}
    \label{tab:integrity2}
\end{table}


\begin{table}[!htb]
    \centering
\begin{tabular}{ccc}
\toprule
 $g$ & acc($\mathcal{D}, g$) & acc($\mathcal{D}^{*}_t, g$) \\
\midrule
ResNet34 & $84.0 \pm 7.5$ & $14.0 \pm 3.2$ \\

MobileNetV2 & $77.0 \pm 6.5$ & $12.7 \pm 2.9$ \\

VGG11 & $83.1 \pm 1.8$ & $13.3 \pm 1.5$\\
\bottomrule
\end{tabular}
    \caption{Results of experiments on the integrity of the method, part $3$. The training dataset $\mathcal{D}$ is SVHN. }
    \label{tab:integrity3}
\end{table}


\section{Conclusion}
\label{s:conclusion}
In this paper, we propose a novel trigger set-based watermarking approach to address intellectual property protection in the context of black-box model stealing attacks. Our method produces trigger sets that are transferable between the source model and the surrogate models with high probability. Our approach is model-agnostic, does not require any additional model training, and does not imply any limitations on the size of the trigger set. Thus, it can be applied to any model without causing performance sacrifice and minimal computational overhead for trigger set generation. We evaluate our approach on multiple benchmarks against the concurrent methods and show that our method outperforms state-of-the-art watermarking techniques in all considered experimental setups. Future work includes analysis of the guarantees of transferability of digital watermarks and study of provable integrity for certified ownership verification.

\section*{Acknowledgements}
The authors would like to thank Anastasia Antsiferova from the video group of MSU Graphics and Media Laboratory, Nikita Kotelevskii and Ekaterina Kuzmina from Skolkovo Institute of Science and Technology for useful discussions during the preparation of this paper. This work was partially supported by a grant for research centers in the field of artificial intelligence, provided by the Analytical Center in accordance with the subsidy agreement (agreement identifier 000000D730321P5Q0002) and the agreement with the Ivannikov Institute for System Programming of the Russian Academy of Sciences dated November 2, 2021 No. 70-2021-00142.

\bibliographystyle{named}
\bibliography{ijcai24}

\end{document}


\maketitle

\appendix

\section{Additional Experiments}
In this section, we provide the results of experiments with different threat models and the integrity of our approach.

\subsection{Additional Threat Models}

\subsubsection{Pruning}
We evaluate our watermarking approach against pruning threat model. The source model is modified by pruning its less active neurons, possibly leading to the removal of the watermarks embedded into it. In our experiments, we apply pruning to the source model and evaluate its performance and watermark accuracy change. For each experiment, we perform pruning attack $N_i=3$ times to obtain surrogate models $f^{*}$. In  Table \ref{tab:pruning} and Figures  \ref{fig:enter-label}-\ref{fig:enter-label2}, we report the results of our evaluation. It should be mentioned that a pruning attack can not effectively remove the watermarks without leading to a significant performance decrease. Thus, it can be seen that our approach can be effectively used against the pruning threat model. 
\begin{figure}
    \centering
    \includegraphics[width=1\linewidth]{Figs_PDF/cifar10_prun (4).pdf}
    \caption{Pruning attack results. Training dataset $\mathcal{D}$ is CIFAR-10, architecture of $f^{*}$ is ResNet34.}
    \label{fig:enter-label}
    \centering
    \includegraphics[width=1\linewidth]{Figs_PDF/cifar100_prun (4).pdf}
    \caption{Pruning attack results. Training dataset $\mathcal{D}$ is CIFAR-100, architecture of $f^{*}$ is ResNet34.}
    \label{fig:enter-label2}
\end{figure}

\input{FormattingGuidelines-IJCAI-24/pruning_table}

\subsubsection{Fine-tuning}

It was reported that fine-tuning may also lead to watermark removal \cite{kim2023margin}. To evaluate our method against this threat model, we fine-tune the source network $f$ for $100$ epochs with learning rate of $0.1$ and obtain $N_i=3$ surrogate models $f^{*}.$ In Table \ref{tab:finetune}, we report the results of our experiments. Although the trigger set accuracy decreases from the initial value of  $100\%$, it still remains high enough for the ownership claim.  

\input{FormattingGuidelines-IJCAI-24/finetune}

\subsection{Integrity of the Method}
To satisfy the integrity property, the method has to be able to distinguish between stolen models and independent (not stolen) models $g$. To check this property, we evaluate  our approach on independent models $g$, which were trained in several different setups: 

\begin{itemize}
    \item In the first setting, we train $N_i = 64$ models on random subsets of initial training dataset. Architectures of independent models are either ResNet34 or WideResNet28 \cite{zagoruyko2016wide}. For each model, the training dataset is twice as small as the dataset of the source model. We report the results of this setting in Table \ref{tab:integrity1}.

    \item In the second setting, we evaluate models of different architectures trained on the whole CIFAR-10 dataset. We consider VGG \cite{DBLP:journals/corr/SimonyanZ14a}, ShuffleNetV2 \cite{ma2018shufflenet}, ResNet20, ResNet32, ResNet44, ResNet56 \cite{he2016deep}, RepVGG \cite{ding2021repvgg} and MobileNetV2 \cite{sandler2018mobilenetv2} architectures. We report the results of this setting in Table \ref{tab:integrity2}.

    \item In the third setting, we train $N_i=3$ models of different architectures on different dataset, $\mathcal{D}=$ SVHN \cite{netzer2011reading}. All the models were trained for $30$ epochs with SGD optimizer, learning rate of $0.1$ and weight decay of $10^{-4}$. We report the results of this setting in Table \ref{tab:integrity3}. 
\end{itemize}
It is expected that if the model is independent, then it is less similar to the source model and, hence, its accuracy on the trigger set is lower than that of the source model and stolen models.

For all the experiments, we use the verification procedure from the Discussion section of the manuscript. As the results, we report the accuracy of independent models on respective training datasets $\mathcal{D}$ and accuracy on watermarks $\mathcal{D}^{*}$.

Notably, the independent models from the setups below have different degrees of similarity with the source model.  Regardless of architecture, the model trained on the same dataset as the source model partially retains the behavior of the source model on the watermarks. However, trigger set accuracy is significantly lower then the one of surrogate (stolen) models. Our experiments show that the least similar independent models are the ones trained on the different dataset, regardless of architecture. 

To sum up, we want to outline that our model can to catch the difference between stolen model and independent models; the larger the dissimilarity between the independent model and the source model, the greater the difference in behavior on the trigger set.

\begin{table}[t]
    \centering
\begin{tabular}{cccc}
\toprule
 $g$ & acc($\mathcal{D}, f$) & acc($\mathcal{D}, g$) & acc($\mathcal{D}^{*}, g$) \\
\midrule
ResNet34 & \multirow{2}{*}{$90.6 \pm 0.4$} & $77.9 \pm 2.6$ & $55.7 \pm 7.2$ \\

WideResNet28 &  & $92.3 \pm 0.3$ & $53.9 \pm 5.0$ \\

\bottomrule
\end{tabular}
    \caption{Results of experiments on the integrity of the method, part $1$. The training dataset $\mathcal{D}$ is CIFAR-10.}
    \label{tab:integrity1}
\end{table}

\begin{table}[t]
    \centering
\begin{tabular}{cccc}
\toprule
 $g$ & acc($\mathcal{D}, f$) & acc($\mathcal{D}, g$) & acc($\mathcal{D}^{*}, g$) \\
\midrule
VGG & \multirow{5}{*}{$90.6 \pm 0.4$} & $93.7 \pm 0.5$ & $58.2 \pm 3.9$ \\
 
 ShuffleNetV2 & & $92.6 \pm 1.5$ & $57.3 \pm 2.7$\\
 
ResNet &  & $93.6 \pm 0.7$ & $53.9 \pm 5.0$ \\

RepVGG & & $94.8 \pm 0.3$ & $53.3 \pm 2.4$\\
MobileNetV2 & & $93.6 \pm 0.5$ & $57.0 \pm 2.2$\\

\bottomrule
\end{tabular}
    \caption{Results of experiments on the integrity of the method, part $2$. The training dataset $\mathcal{D}$ is CIFAR-10.}
    \label{tab:integrity2}
\end{table}


\begin{table}[!htb]
    \centering
\begin{tabular}{ccc}
\toprule
 $g$ & acc($\mathcal{D}, g$) & acc($\mathcal{D}^{*}, g$) \\
\midrule
ResNet34 & $84.0 \pm 7.5$ & $14.0 \pm 3.2$ \\

MobileNetV2 & $77.0 \pm 6.5$ & $12.7 \pm 2.9$ \\

VGG11 & $83.1 \pm 1.8$ & $13.3 \pm 1.5$\\
\bottomrule
\end{tabular}
    \caption{Results of experiments on the integrity of the method, part $3$. The training dataset $\mathcal{D}$ is SVHN. }
    \label{tab:integrity3}
\end{table}

\bibliographystyle{named}
\bibliography{ijcai24}

%% file: FormattingGuidelines-IJCAI-24/table_with_diffs.tex
\begin{table}[t]
    \centering
\begin{tabular}{cccc}
\toprule
 $\mathcal{D}$ & $\hat{\mathcal{D}}$ & Attack & $\Delta$ \\
\midrule
 \multirow{3}{*}{CIFAR-10} & \multirow{3}{*}{CIFAR-10} & Soft & $44.65 \pm 0.04$ \\
 & & Hard & $54.79 \pm 1.10$ \\
 & & RGT & $49.61 \pm 0.36$ \\ 
 \midrule 
\multirow{3}{*}{CIFAR-100} & \multirow{3}{*}{CIFAR-100} & Soft & $82.03 \pm 1.91$ \\
& & Hard & $82.21 \pm 3.52$ \\
& & RGT & $82.29 \pm 2.92$ \\ 
\midrule  
CIFAR-10 & SVHN & Soft & $46.33 \pm 0.07$ \\

\bottomrule
\end{tabular}
    \caption{The $l_2-$norm of the difference $\Delta$ of models' parameters between the source model $f$ and surrogate models $f^{*}$ for different types of stealing attacks. The architecture of the source model and surrogate models is ResNet34.}
    \label{tab:table_diffs}
\end{table}

%% file: FormattingGuidelines-IJCAI-24/table1.tex
\begin{table*}[htb]
    \centering
\begin{tabular}{lccccc}
\toprule \multirow{2}{*}{ Method } & \multirow{2}{*}{ Metric } & \multirow{2}{*}{ Source model $f$} & \multicolumn{3}{c}{ Surrogate models $f^{*}$} \\
& & & Soft-label & Hard-label & RGT \\
\midrule

EWE \cite{jia2021entangled} & \multirow{4}{*}{CIFAR-10 acc. (\%) }  & $86.10 \pm 0.54$ & $83.97 \pm 1.02$ & $82.22 \pm 0.50$ & $88.88 \pm 0.35$ \\
RS \cite{bansal2022certified} & & $84.17 \pm 1.01$ & $88.93 \pm 1.18$ & $89.62 \pm 0.97$ & $90.14 \pm 0.08$ \\
MB \cite{kim2023margin} & & $87.81 \pm 0.76$ & $91.17 \pm 0.76$ & $91.88 \pm 0.40$ & $93.05 \pm 0.20$\\
\textbf{Probabilistic (Ours)} & & $\textbf{91.00} \pm \textbf{0.00}$ & $\textbf{92.60} \pm \textbf{0.91}$ & $\textbf{94.87} \pm \textbf{0.59}$ & $\textbf{99.42} \pm \textbf{0.02}$ \\
\midrule

EWE \cite{jia2021entangled} & \multirow{4}{*}{Trigger set acc. (\%) }  & $26.88 \pm 8.32$ & $51.01 \pm 5.58$ & $36.05 \pm 6.48$ & $1.64 \pm 1.05$ \\
RS \cite{bansal2022certified} & & $95.67 \pm 4.93$ & $7.67 \pm 4.04$ & $6.33 \pm 1.15$ & $3.00 \pm 0.00$ \\
MB \cite{kim2023margin} & & $100.00 \pm 0.00$ & $82.00 \pm 1.00$ & $51.33 \pm 4.93$ & $72.67 \pm 6.66$\\
\textbf{Probabilistic (Ours)} & & $100.00 \pm 0.00$ & $\textbf{85.10} \pm \textbf{6.33}$ & $\textbf{73.70} \pm \textbf{4.65}$ & $\textbf{78.00} \pm \textbf{5.58}$ \\
\midrule

EWE \cite{jia2021entangled} & \multirow{4}{*}{CIFAR-100 acc. (\%) }  & $55.11 \pm 1.67$ & $53.00 \pm 1.57$ & $46.78 \pm 1.00$ & $63.73 \pm 0.40$ \\
RS \cite{bansal2022certified} & & $59.87 \pm 2.78$ & $65.66 \pm 1.53$ & $65.79 \pm 0.39$ & $64.99 \pm 0.30$ \\
MB \cite{kim2023margin} & & $62.13 \pm 4.36$ & $\textbf{67.66} \pm \textbf{0.36}$ & $\textbf{70.65} \pm \textbf{0.49}$ & $\textbf{70.24} \pm \textbf{0.46}$\\
\textbf{Probabilistic (Ours)} & & $\textbf{66.70} \pm \textbf{0.00}$ & $67.49 \pm 0.03$ & $68.05 \pm 0.73$ & $67.85 \pm 0.04$ \\
\midrule

EWE \cite{jia2021entangled} & \multirow{4}{*}{Trigger set acc. (\%) }  & $68.14 \pm 10.16$ & $30.90 \pm 11.34$ & $15.10 \pm 5.64$ & $5.73 \pm 3.42$ \\
RS \cite{bansal2022certified} & & $99.00 \pm 1.00$ & $2.67 \pm 1.53$ & $4.33 \pm 4.16$ & $2.00 \pm 1.00$ \\
MB \cite{kim2023margin} & & $100.00 \pm 0.00$ & $70.67 \pm 7.57$ & $40.00 \pm 8.89$ & $62.66 \pm 10.12$\\
\textbf{Probabilistic (Ours)} & & ${100.00} \pm {0.00}$ & $\textbf{78.80} \pm \textbf{2.93}$ & $\textbf{74.70} \pm \textbf{3.16}$ & $\textbf{79.10} \pm \textbf{2.77}$ \\
\bottomrule

\end{tabular}
    \caption{Watermarking performance is reported against functionality stealing methods. The best performance is highlighted in bold. It can be seen that our approach outperforms the other methods of ownership verification by a notable margin.}
    \label{tab:main_table}
\end{table*}

%% file: FormattingGuidelines-IJCAI-24/table_additional_exp.tex
\begin{table*}[t]
    \centering
\begin{tabular}{ccccccc}
\toprule
Method & 
$f^{*}$& 
$\hat{\mathcal{D}}$ & acc$(\mathcal{D}, f)$ & acc$({\mathcal{D}}^{*}, f)$ & acc$(\mathcal{D}, f^{*})$ & acc$({\mathcal{D}}^{*}, f^{*})$\\
\midrule
\multirow{2}{*}{MB \cite{kim2023margin}} & ResNet34 & SVHN & $87.81 \pm 0.76$ & $100.0\pm 0.00$ & $63.99 \pm 3.90$ & $72.00 \pm 6.08$\\
& VGG11 & CIFAR-10 & $87.81 \pm 0.76$ & $100.0\pm 0.00$ & $86.00 \pm 2.17$ & $32.00 \pm 7.21$ \\
\midrule 
\multirow{2}{*}{Probabilistic (ours)} & ResNet34 & SVHN & $91.00 \pm 0.00$ & $100.0\pm 0.00$ & $73.01 \pm 1.18$ & $77.70 \pm 2.90$ \\
& VGG11 & CIFAR-10 & $91.00 \pm 0.00$ & $100.0\pm 0.00$ & $89.24 \pm 2.69$ & $80.10 \pm 3.86$\\
\bottomrule
\end{tabular}
    \caption{Results of watermarking approaches in the setting when either the training dataset or source model's architecture is unknown to the adversary. Our approach outperforms the baseline in terms of the initial accuracy of the source model and the trigger set accuracy of surrogate models. }
    \label{tab:table_additional}
\end{table*}

%% file: FormattingGuidelines-IJCAI-24/table_with_parameters.tex
\begin{table*}[ht]
    \centering
\begin{tabular}{cccccc}
\toprule 

Surrogate model $f^{*}$ & Training dataset $\mathcal{D}$ & Surrogate dataset $\hat{\mathcal{D}}$& Stealing method & Parameter $m$ & Parameter $\delta$ \\ 
\midrule

\multirow{7}{*}{ResNet34} & \multirow{4}{*}{CIFAR-10} & \multirow{3}{*}{CIFAR-10} & Soft-label & $64$ & $40.0$\\
& & & Hard-label & $64$ & $40.0$\\
& & & RGT & $256$ & $40.0$\\
\cmidrule{4-6}
& & SVHN & Soft-label & $256$ & $40.0$\\
\cmidrule{4-6}
& \multirow{3}{*}{CIFAR-100} & \multirow{3}{*}{CIFAR-100} & Soft-label & $64$ & $40.0$\\ 
& & & Hard-label & $64$ & $40.0$ \\
& & & RGT & $64$ & $40.0$ \\
\midrule
VGG11 & CIFAR-10 & CIFAR-10 & Plain training & $64$ & $40.0$\\
\bottomrule
\end{tabular}
    \caption{Values of hyperparameters used in each experiment. For all the experiments, the size $n$ of the trigger set and performance threshold $\tau$ are $n=100$ and $\tau=1.0$. For the RGT stealing method, the regularization coefficient from Eq.~\eqref{eq:rgt_loss} is set to be $\gamma=0.3$.}
    \label{tab:parameters_table}
\end{table*}

%% file: FormattingGuidelines-IJCAI-24/pruning_table.tex
\begin{table}[ht]
    \centering
\scalebox{1.0}{\begin{tabular}{cccc}
\toprule
$\mathcal{D}$ & Pruning ratio & acc($\mathcal{D}, f^{*}$) & acc($\mathcal{D}^{*}, f^{*}$) \\
\midrule
\multirow{8}{*}{CIFAR-10} & $0.1$ & $90.6 \pm 0.4 $ &  $86.1 \pm 10.0$ \\

& $0.2$ & $90.6 \pm 0.4 $  & $86.1 \pm 9.9$ \\

& $0.3$ & $90.6 \pm 0.4 $ &  $81.0 \pm 13.1$ \\

& $0.4$ & $89.6 \pm 0.4 $ &  $84.0 \pm 10.5$ \\

&$0.5$ & $87.0 \pm 1.0 $ &  $83.9 \pm 7.0$ \\

& $0.6$ &  $84.0 \pm 2.1 $ & $66.1 \pm 5.2$ \\
&$0.7$ &  $70.6 \pm 4.9$ & $45.3 \pm 9.2$ \\

&$0.8$ &  $35.3 \pm 11.7$ & $32.4 \pm 10.4$ \\

\midrule

\multirow{8}{*}{CIFAR-100} & $0.1$ & $67.0 \pm 0.0$ & $85.0 \pm 10.1$ \\
& $0.2$ &$66.3 \pm 0.1$  & $87.1 \pm 9.2$ \\
& $0.3$ &  $66.0 \pm 0.0$ & $86.3 \pm 10.2$ \\
& $0.4$ & $64.0 \pm 0.0$  & $84.2 \pm 10.0$ \\
&$0.5$ & $59.0 \pm 0.0$  & $77.0 \pm 9.7$ \\
& $0.6$ &  $45.6 \pm 1.7$ & $50.2 \pm 5.3$ \\
&$0.7$ & $23.3 \pm 3.3$  & $22.1 \pm 7.0$ \\
&$0.8$ &  $4.3 \pm 2.6$ & $3.2 \pm 1.0$ \\

\bottomrule
\end{tabular}}
    \caption{Results of experiments against pruning attack. It is notable that pruning can not effectively remove the watermarks without leading to significant performance decrease.}
    \label{tab:pruning}
\end{table}

%% file: FormattingGuidelines-IJCAI-24/finetune.tex
\begin{table}[t]
    \centering
\begin{tabular}{cccc}
\toprule
 $\mathcal{D}$ & acc($\mathcal{D}, f$) & acc($\mathcal{D}, f^{*}$) & acc($\mathcal{D}^{*}, f^{*}$) \\
\midrule
{CIFAR-10} & $90.6 \pm 0.4$ & $92.1 \pm 0.2$ & $74.3 \pm 2.9$ \\
 \midrule 
{CIFAR-100} & $67.0 \pm 0.0$ & $70.3 \pm 0.2$ & $84.1 \pm 7.0$ \\

\bottomrule
\end{tabular}
    \caption{Results of experiments against fine-tuning. Notably, the trigger set accuracy decreases for both experiments. Nevertheless, it is still reasonably high. }
    \label{tab:finetune}
\end{table}